\documentclass[authoryear,3p,two column,times,final]{elsarticle}
\usepackage{aas_macros}

\usepackage{setspace}

\usepackage{bm}
\usepackage{array}
\usepackage{amssymb}
\usepackage{amsmath}
\usepackage[usenames,dvipsnames]{xcolor}
\usepackage{overpic}
\usepackage{graphics}
\usepackage{epstopdf}
\usepackage[font=footnotesize]{subcaption}
\usepackage{afterpage}
\usepackage{pdfpages}

\usepackage[backref,breaklinks,colorlinks,citecolor=blue]{hyperref}        
\usepackage[all]{hypcap}

\usepackage{fancyhdr}
\newcommand{\mearth}{M$_{\oplus}$\,}

\journal{Earth and Planetary Science Letters}

\begin{document}

\begin{frontmatter}

\title{Collisional stripping of planetary crusts}

\author[bphys,dav]{Philip J. Carter\corref{cor}}
\ead{pjcarter@ucdavis.edu}

\author[bphys]{Zo\"{e} M. Leinhardt}
\author[bes]{Tim Elliott}
\author[dav]{Sarah T. Stewart}
\author[bes]{Michael J. Walter}

\cortext[cor]{Corresponding author}

\address[bphys]{School of Physics, University of Bristol, H. H. Wills Physics Laboratory, Tyndall Avenue, Bristol BS8 1TL, UK}
\address[dav]{Department of Earth and Planetary Sciences, University of California Davis, One Shields Avenue, Davis, CA 95616, USA}
\address[bes]{School of Earth Sciences, University of Bristol, Wills Memorial Building, Queen's Road, Bristol BS8 1RJ, UK}


\begin{abstract}
Geochemical studies of planetary accretion and evolution have invoked various degrees of collisional erosion to explain differences in bulk composition between planets and chondrites.
Here we undertake a full, dynamical evaluation of `crustal stripping' during accretion and its key geochemical consequences.
Crusts are expected to contain a significant fraction of planetary budgets of incompatible elements, which include the major heat producing nuclides. 
We present smoothed particle hydrodynamics simulations of collisions between differentiated rocky planetesimals and planetary embryos.  We find that the crust is preferentially lost relative to the mantle during impacts, and we have developed a scaling law based on these simulations that approximates the mass of crust that remains in the largest remnant. Using this scaling law and a recent set of $N$-body simulations of terrestrial planet formation, we have estimated the maximum effect of crustal stripping on incompatible element abundances during the accretion of planetary embryos. 
We find that on average approximately one third of the initial crust is stripped from embryos as they accrete, which leads to a reduction of $\sim$20\% in the budgets of the heat producing elements if the stripped crust does not reaccrete.
Erosion of crusts can lead to non-chondritic ratios of incompatible elements, but the magnitude of this effect depends sensitively on the details of the crust-forming melting process on the planetesimals.
The Lu/Hf system is fractionated for a wide range of crustal formation scenarios.  Using eucrites (the products of planetesimal silicate melting, thought to represent the crust of Vesta) as a guide to the Lu/Hf of planetesimal crust partially lost during accretion, we predict the Earth could evolve to a superchondritic $^{176}$Hf/$^{177}$Hf (3-5 parts per ten thousand) at present day. Such values are in keeping with compositional estimates of the bulk Earth. 
Stripping of planetary crusts during accretion can lead to detectable changes in bulk composition of lithophile elements, but the fractionation is relatively subtle, and sensitive to the efficiency of reaccretion.

\end{abstract}


\end{frontmatter}


\section{Introduction}

Terrestrial planets are thought to form from the accumulation of planetesimals into planetary embryos, followed by a period in which these embryos undergo giant impacts to form a final system of planets \citep[e.g.][]{Morbi12}. Accretion of planetesimals onto larger bodies is a key process during both intermediate and late stages of planet formation, but this process is not one of monotonic growth. The collisions experienced by the planets as they grow can also lead to loss of material and the final composition of a planetary body can be influenced by preferential erosion of chemically distinct layers \citep[e.g.][]{Carter15}.

Crusts are expected to contain a significant fraction of planetary incompatible element budgets (elements that are concentrated during magmatic processes because they more strongly partition into the melt than residual solid during partial melting), and it has previously been suggested that preferential removal of the outer crustal layers of planetary building blocks alter bulk compositions \citep[e.g.][]{Greenwood05,ONeill08,Boujibar15,Jellinek+Jackson15}. Since there is ample evidence of the early differentiation of planetesimals  \citep[e.g.][]{Lugmair98,Srinivasan99,Bizzarro05,Amelin08,Kruijer14}, collisional erosion is an obvious means of removing the chemically distinct outer layers of growing planets.  Several studies have previously demonstrated that collisions during accretion can influence the major element compositions of the resulting objects (e.g.\ \citealt{Marcus09,Marcus10,Bonsor15,Carter15,Dwyer15}), but most of the work on individual collision outcomes has not focused on heterogeneous colliders \citep[e.g.][]{Leinhardt12}, or has considered only giant impacts \citep{Marcus09,Marcus10}.  Moreover these studies have not quantified the budgets of important trace elements\footnote{Trace elements are defined as having concentrations $<$1000\,ppm} used to characterise planetary bodies.

The modification of planetary compositions during growth has consequences for the long-standing paradigm 
that primitive meteorites, chondrites, represent the building blocks of planets. Chondritic ratios of cosmochemically refractory elements (elements with high condensation temperatures in a hydrogen-rich environment), such as the rare earth elements, have been taken to provide a robust bulk planetary reference.  Yet, some doubt has been cast on the reliability of this approach given the possibility that the Earth has non-chondritic $^{142}$Nd/$^{144}$Nd \citep{BoyetandCarlson05}. Proposed means to account for the terrestrial $^{142}$Nd/$^{144}$Nd include: the Earth is made from chondrites different to those in our current collection \citep[e.g.][]{Huang13,Burkhardt16}, the Earth's mantle is incompletely sampled \citep[e.g.][]{BoyetandCarlson05,Labrosse07}, or the Earth accreted from differentiated bodies that had experienced prior collisional erosion \citep[e.g.][]{Bourdon08,ONeill08}.
Since $^{142}$Nd is the daughter of the extinct radioisotope $^{146}$Sm ({half life $\sim$100\,Myr, \citealt{Meissner87}}), 
a reservoir enriched in Sm relative to Nd (i.e.\ incompatible element depleted) early in the evolution of the solar system (within the first 100\,Myr) would evolve into a reservoir with elevated $^{142}$Nd/$^{144}$Nd.
Whilst the idea that the Earth accreted from distinct material has recently gained significant traction \citep[see also][]{Bouvier16} as a preferred explanation for super-chondritic terrestrial $^{142}$Nd/$^{144}$Nd, recent dynamical models have underscored the potential for collisional modification of bulk planetary compositions during accretion \citep[e.g.][]{Bonsor15,Carter15}, at least for readily fractionated materials. Here we specifically investigate the importance of such collisional losses from differentiated bodies for the ratios of key refractory, lithophile element isotopes (Sm-Nd, Lu-Hf). 

\subsection{Previous work}

\citet{Marcus09} presented simulations of giant impacts onto Earth and super-Earth sized bodies using differentiated impactors with Earth-like composition and masses in the range 0.25--10\,\mearth {(where \mearth is the mass of the Earth)}. As well as examining the catastrophic disruption criteria, they derived a scaling law for the change in bulk composition due to mantle stripping, demonstrating that the iron-to-silicate ratio of the largest remnant increases with increasing impact energy. That work was extended to ice-rock planets in \citet{Marcus10}.

\citet{Leinhardt12} used the results of a variety of previous impact simulations (including  \citealt{Marcus09}) and conducted new $N$-body simulations of smaller (homogeneous) planetesimal collisions, using 10\,km radius targets, to determine the effects of collision parameters on collision outcome. They derived scaling laws to describe the size and velocity distribution of the collision remnants (see also \citealt{Leinhardt15}). This work demonstrated that the collision outcome is velocity dependent, allowing correction for the impact angle and mass ratio.

We have previously applied the \citet{Marcus09} mantle stripping law and the state-of-the-art collision model from \citet{Leinhardt12} and \citet{Leinhardt15} to examine the changes in bulk composition during accretion of planetary embryos using high resolution $N$-body simulations \citep{Bonsor15,Carter15}. These simulations of the intermediate stages of planet formation, from $\sim$100$\,$km planetesimals to planetary embryos, explored two contrasting dynamical scenarios: a calm disc unperturbed by giant planets, and the dynamically hot Grand Tack model \citep{Walsh11}. We found that significant compositional change can occur, in terms of core/mantle ratio, in growing planetesimals, especially in dynamically hot protoplanetary discs \citep{Carter15}.

{Here we explore the effects of collisions on the crusts of planetesimals by conducting a new set of impact simulations, and explore the implications for planetary compositions by applying scaling laws to previous $N$-body simulations of planet formation.} We begin by describing the impact simulations, and the post-processing of $N$-body simulations to examine crust stripping during accretion (section \ref{s:method}). We present the results of the impact simulations in section \ref{s:SPHresults} and the results of the post-processing in section \ref{s:postprocess}, then discuss these results in section \ref{s:discussion}.


\section{Numerical Methods}\label{s:method}

\subsection{Hydrodynamic modelling}

We simulated collisions between planetesimals and embryos using a version of the GADGET-2 smoothed particle hydrodynamics (SPH) code \citep{Springel05} modified for planetary collision calculations. This modified version of GADGET-2 calculates thermodynamic quantities by interpolating tabulated equations of state, for details see \citet{Marcus09}, \citet{Marcus10}, and \citet{Cuk12}. As in previous studies \citep{Marcus09,Cuk12,Lock17}, we modelled the silicate layers and metallic cores of terrestrial planet embryos as pure forsterite and pure iron using tabulated equations of state obtained from the M-ANEOS model (\citealt{Melosh07}; these tables are available in the online supplement to \citealt{Cuk12}).

The initial differentiated planetesimals were comprised of 22\,wt\% iron and 78\,wt\% forsterite (this smaller iron core compared to \citet{Marcus09} is consistent with chondritic building blocks). 
The bodies were initialised with a simple temperature profile approximating the state of planetesimals in the first few million years of evolution. 
Since the planetesimals are differentiated they should have undergone large-scale melting, and so, based on the melting model from \citet{Hevey06}, we set the temperature to 1850\,K 
from the centre to $0.8R$ (where $R$ is the radius of the body), dropping linearly to 300\,K at the surface.

\begin{figure}
\centering
\includegraphics[width=0.48\textwidth]{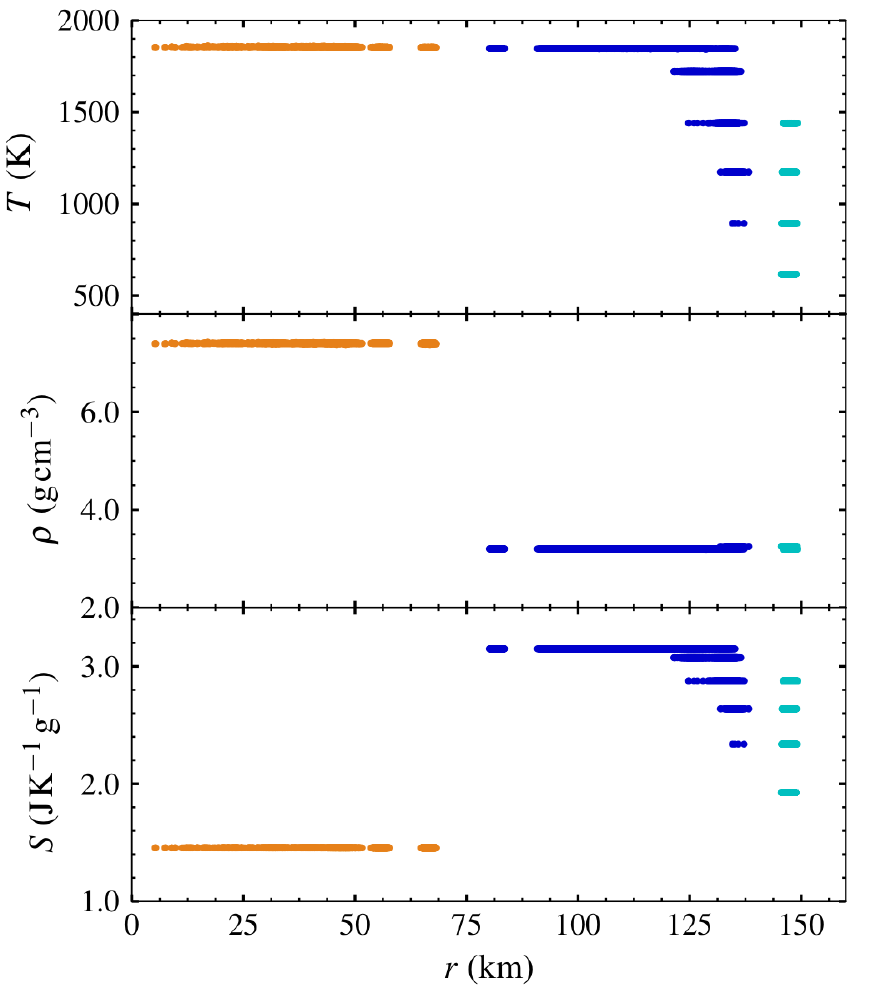}
\caption{Radial temperature, density and specific entropy structure of a $10^{-5}\,$\mearth  planetesimal after gravitational equilibration. The orange and blue points represent particles in the iron core, and forsterite mantle respectively, the cyan points indicate the crust-tracing silicate particles. \label{f:initialcond}}
\end{figure}

The bodies were then evolved through two equilibration steps each lasting 10 hours. Because the initial density and temperature distribution are not perfectly in gravitational equilibrium, large particle motions are induced as the body evolves under self-gravity. Thus, to accelerate convergence to equilibrium, particle velocities are damped by a factor of 50\% per time step during the first 10 hour period. Without damping, these particle velocity perturbations can generate strong shocks that disturb the desired thermal profile (sometimes in a dramatic fashion).
In the second equilibration phase the body evolves under self-gravitational forces (without additional damping) towards a hydrostatic profile.
In many cases particle motions during equilibration led to a single particle reaching escape velocity.
The state of a planetesimal of mass $1 \times 10^{-5}\,$\mearth after equilibration is shown in Figure \ref{f:initialcond}. This equilibration procedure does not significantly alter the temperature profile of the body.

We used target masses of $1 \times 10^{-5}\,$\mearth, $5 \times 10^{-5}\,$\mearth, $1 \times 10^{-3}\,$\mearth, and $1 \times 10^{-1}\,$\mearth, mostly comprised of  $2 \times 10^{4}$ particles; more extreme mass ratios require higher resolution in order for the projectile to be well resolved. For projectile/target mass ratios smaller than 0.1 we used targets comprised of $2 \times 10^{5}$ particles. A subset of collisions were run with ten times as many particles to check sensitivity to resolution, producing equivalent results.

We explored mass ratios of 0.01, 0.02, 0.05, 0.1, 0.2, 0.3, 0.4, 0.5 and 1.0, with the projectile having the same mass per particle as the target. Impact parameters of 0.1, 0.25, 0.4, 0.5, and 0.8 were selected to provide good coverage of the parameter space of collisions that occurred in the simulations of \citet{Carter15}. Impact velocities ranged from 0.8 to 15 times the mutual escape velocity\footnote{$\,v_\textrm{esc}=\sqrt{2GM_{\mathrm{Tot}} / R}$, where $M_{\mathrm{Tot}}$ is the sum of the target and projectile masses, and $R$ is the radius of a sphere of mass $M_{\mathrm{Tot}}$ with the same bulk density as the target  \citep{Leinhardt12}.}, $v_\textrm{esc}$. The full list of simulations is available in the supplementary material (Tables S1 and S2).

In addition to the separate iron core and forsterite mantle, the outer silicate layer of each body was tagged to represent the planetesimal's crust (see Figures \ref{f:initialcond} and \ref{f:collseq}). Although this material is labelled as crust, it is compositionally identical to the mantle for the dynamic calculations and we assume that the small differences in densities of real planetary crusts have minimal effect. This outer layer of SPH particles (4980--5370 particles for the standard resolution targets) may not fully resolve the crust, depending on the unknown thickness of the crust on such bodies, and so we assume some fraction of each surface particle represents crustal material \citep[as in][]{Marinova08} and scale the crust mass accordingly. Ascribing distinct chemical characteristics to crustal material is only done in the post-processing stage (see section \ref{s:crustcomp}). 

Bodies were initially separated by an additional projectile radius (i.e. the separation between the body centres in the direction of motion was set to $R_\mathrm{targ}+2R_\mathrm{proj}$) such that they were not already in contact at the start of the run. Most simulations were run for at least 24 hours {(model time within the simulation)} to allow reaccumulation of debris onto the largest remnant. 

The mass of the largest remnant following each collision was calculated using the same approach used by \citet{Marcus09}. The potential and kinetic energies of all particles were calculated with respect to the potential minimum, and the centre of mass for all bound particles calculated. This process was repeated on unbound particles until the remnant mass converged. The remnant mass was considered unreliable if it consisted of fewer than 100 particles, in these cases the impact was ignored in further analyses (and no mass is recorded in Table S1). {The same approach was followed to identify the second largest remnant, ignoring particles already identified as bound to the largest remnant.} The mass of the resulting iron core was found simply from the mass of iron particles in the largest remnant. Since the number, and hence mass, of `crust' particles in a body depends on its size, we converted the number of crust particles in the remnant to a mass by assuming the initial bodies each had a crust mass of 15\% of the body's total mass (based on observations of Vesta, see section \ref{s:crustcomp}), and scaling the crust mass fraction of the outer layer particles accordingly.

Finally, the catastrophic disruption threshold, $Q^{\prime *}_{\mathrm{RD}}$ (the specific energy required to disperse half of the total mass), was estimated by fitting a straight line (according to the universal law from \citealt{Leinhardt12}) to the mass of the largest remnant as a function of the specific energy of the impact, $Q_{\mathrm{R}}$ (calculated using the reduced mass), for each group of target mass, mass ratio and impact parameter. Simulations with largest remnants smaller than 0.3 times the total mass were excluded from these calculations, as these may be in the supercatastrophic regime, and would thus deviate from the linear relation. {In the case of hit-and-run impacts the same procedure was followed using the mass of the second largest remnant, $M_\textrm{slr}$, compared to the projectile to calculate $Q^{\dag \prime *}_{\mathrm{RD}}$ (where the $\dag$ is used to indicate the reverse impact onto the projectile, see \citealt{Leinhardt12}).}

\subsection{{Crustal composition evolution}}\label{s:crustcomp}

{In order to examine the consequences of collisions for the compositions of accreting planets, we used the crustal loss scalong law inferred from the SPH simulations (section \ref{s:SPHresults}) to post-process the same set of high resolution $N$-body simulations used to inform the choice of SPH simulations, those from \citet{Carter15}. Each body was assigned initial crust and mantle compositions as described below, and based on the predicted crust and mantle loss (or gain) as a function of specific impact energy the new crust and mantle masses and compositions of bodies were calculated after each collision.} 
In hit-and-run collisions the target is assumed to be unchanged, but stripping of the projectile's crust is calculated in the same way, using the specific energy of the reverse impact (see \citealt{Leinhardt12}). In the disruption regime, for impacts below $Q^{\prime *}_{\mathrm{RD}}$ the second largest remnant is assumed to come from the projectile, above $Q^{\prime *}_{\mathrm{RD}}$ it is assumed both the largest and second largest remnants are fragments of the target, as indicated by the SPH impacts.

{We track the crustal compositions of the elements of interest, namely Sm, Nd, Lu, Hf, Th and U. This suite of incompatible, refractory lithophile elements allows us to track the isotopic evolution of the Sm-Nd and Lu-Hf systems in addition to the fate of major heat-producing elements. We use Vesta as a suitable analogue for intermediate sized planetesimals, and consider three different approaches for setting crustal compositions. a) We calculate a `model' crustal composition based on existing petrological studies of Vesta \citep{Ruzicka97,Mandler13} and appropriate melt-solid partition coefficients ($D$, see Table \ref{t:part}) taken from \citet{Workman+Hart05}. b) We use analyses of eucritic meteorites \citep{Blicherttoft02,Barrat00}, which are believed to be derived from Vesta, to determine the crustal composition.  c) The consequences of more extreme differentiation experienced in larger bodies during later accretion are assessed with a `KREEP-rich' lunar crust. This numerically combines an anorthosite outer layer with the potassium, rare earth element and phosphorous rich (KREEP) residuum of lunar magma ocean crystallisation \citep[e.g.][]{Snyder92}. Our empirical crustal compositions are reported in Table \ref{t:part} as `eucritic' and `KREEP-rich' respectively. For the model and eucritic crusts we assume a silicate melt fraction of 20\% (i.e.\ the crust mass is 20\% of the silicate mass of the body, $\sim$15\% of the total mass), which is appropriate for the estimated fraction of Vesta's eucritic crust \citep[e.g.][]{Mandler13}, and for the KREEP-rich crust we assume the crust is 5.04\% of the silicate mass, consistent with our calculated lunar crust.
Further details of how we derive these composition are given in \ref{a:crustcomp}.
}

For each of these 3 composition models we explore two scenarios of crustal evolution: 1) after each collision any remaining crust on each body is fully mixed with the residual mantle, and a new crust is formed by melting of this new mantle (which includes any mantle or crust accreted from an impactor); the composition of both crust and mantle can change as a result of the collision; 2) any crust remaining is assumed to redistribute across the planetesimal, but no new crust is created to replace that which was lost (both mantle and crust retain their initial composition).

Initially, and in the case of the first evolution scenario each
time new crust is produced, the required mass of crust matching the empirical composition (or containing all of a species left in the body if it is already so depleted that there is not enough left to match the composition) is formed according to mass-balance calculations (eucritic and KREEP-rich crusts), or the crust
composition is set via element partitioning between the mantle and melt according to measured partition coefficients (model crust).

{Since this is a post-processing calculation using $N$-body simulations that did not track crust, it is not possible to follow the long-term (over multiple orbits) fate of stripped crustal material that is initially unbound from the largest remnant, but is still bound to the star. Thus we calculate the end-member scenario in which we neglect the chemical consequences of reaccretion of this material.}


\section{Results}


\subsection{Simulation results}\label{s:SPHresults}

\begin{figure}
\centering
\includegraphics[width=0.48\textwidth]{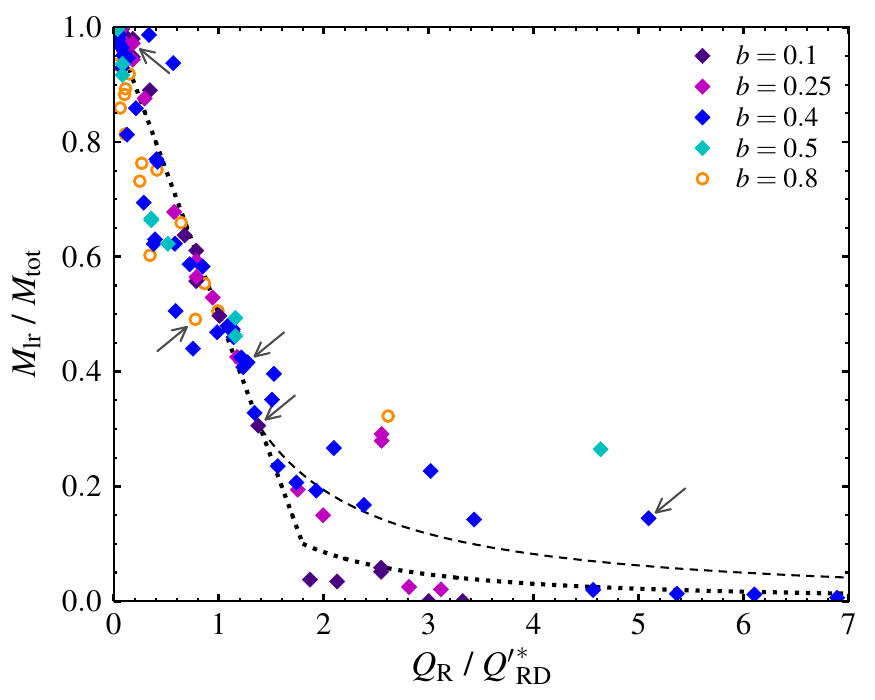}
\caption{Normalised largest remnant mass vs. normalised specific impact energy. The symbol colour denotes the impact parameter. The dotted line is the universal law and, beyond the break in slope at a specific energy of 1.8, the supercatastrophic disruption power law from \citet{Leinhardt12}. The dashed line (also from \citealt{Leinhardt12}) represents laboratory measurements of basalt disruption from \citet{Fujiwara77}. {The open circles are hit-and-run collisions for which $M_\textrm{slr} / M_\textrm{proj}$ vs $Q^\dag_\textrm{R} / Q^{\dag \prime *}_{\mathrm{RD}}$ (see text) is plotted.} The arrows indicate collisions shown in detail in Figure \ref{f:collseq}.\label{f:mlr}}
\end{figure}
Figure \ref{f:mlr} shows the mass of the largest remnant, $M_{\mathrm{lr}}$, against the specific impact energy relative to the catastrophic disruption threshold. The dotted line in this figure represents the `universal law' for the mass of the largest remnant and the power law for the supercatastrophic regime from \citet{Leinhardt12}. Our results show excellent agreement with the \citet{Leinhardt12} model in the accretion and erosion regimes, despite the different methods and impactor sizes, and good agreement with significant scatter in the supercatastrophic regime. Similar results were seen by \citet{Marcus09} (their impact energies were all less than two times the catastrophic disruption threshold). It is worth recalling that the targets simulated in this work fall between the size regimes explored by \citet{Leinhardt12} and \citet{Marcus09}. 

\begin{figure}
\centering
\includegraphics[width=0.48\textwidth]{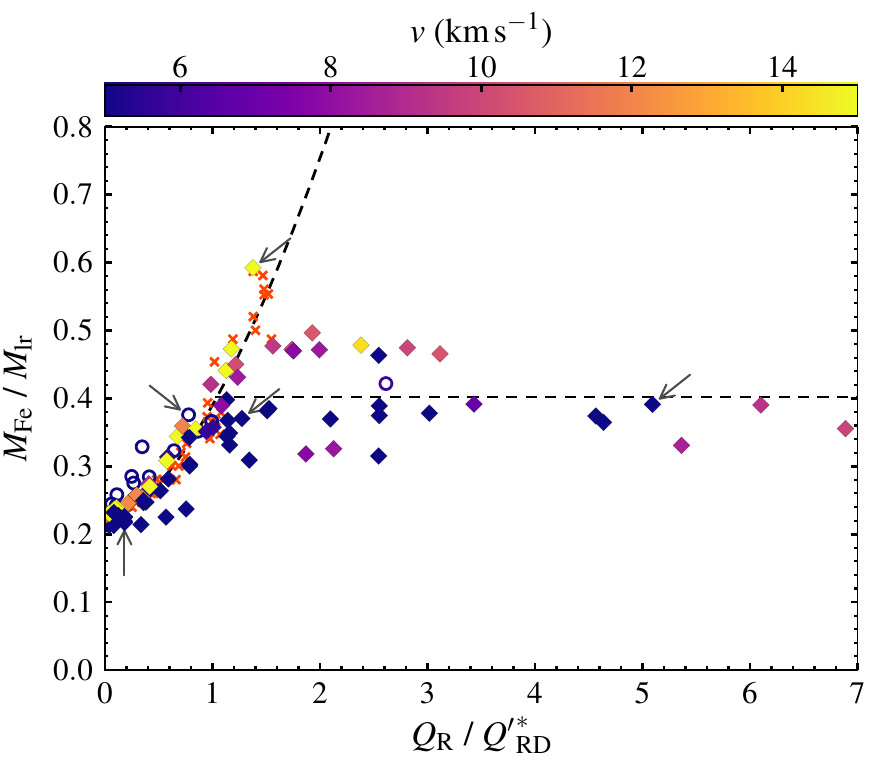}
\caption{Iron fraction for the largest remnant vs. normalised specific impact energy. {Diamonds indicate non-grazing collisions and open circles indicate hit-and-run collisions} from this work, and are coloured according to impact velocity; red crosses (not representative of velocity) are from \citet{Marcus09}, scaled down to match the core fraction used in this work. The dashed curve is a scaled down version of the fit from \citet{Marcus09}. The horizontal dashed line represents a fit to the iron fraction from this study only above a scaled impact energy of 1.0. The arrows indicate collisions shown in detail in Figure \ref{f:collseq}. {For hit-and-run collisions the iron fraction of the second largest remnant and $Q^\dag_\textrm{R} / Q^{\dag \prime *}_{\mathrm{RD}}$ are plotted.} \label{f:corefrac}}
\end{figure}
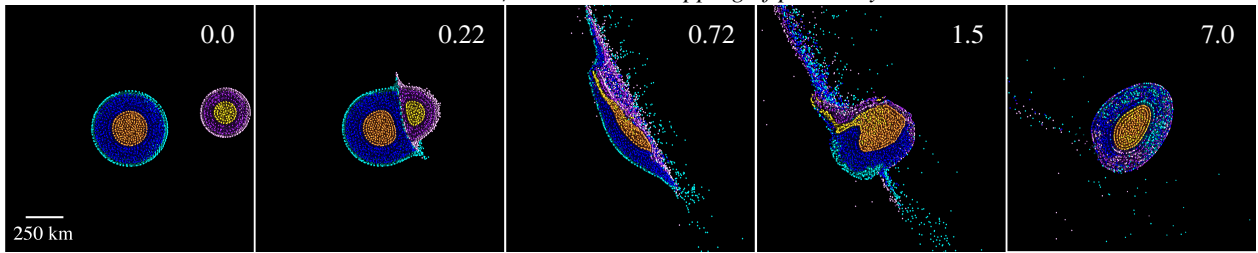
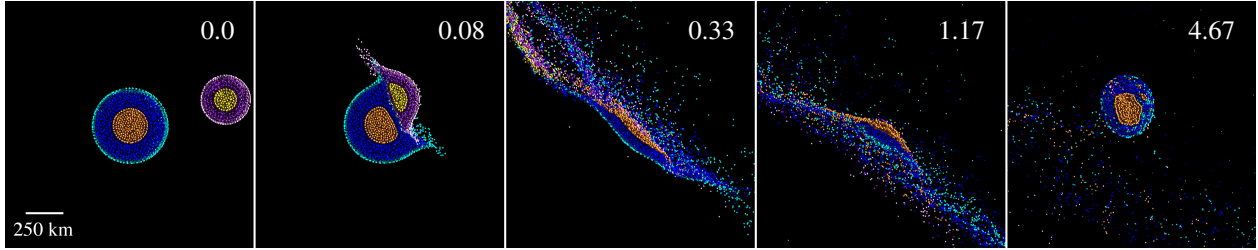
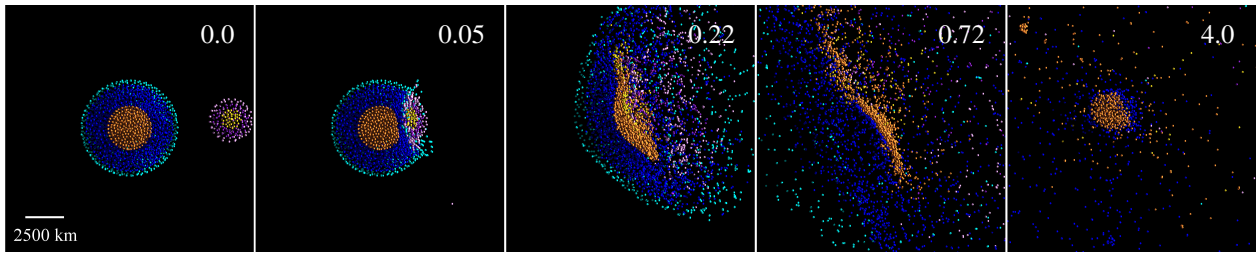
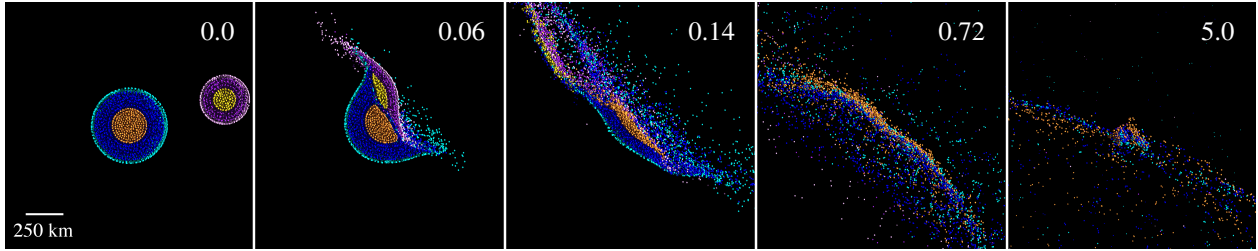
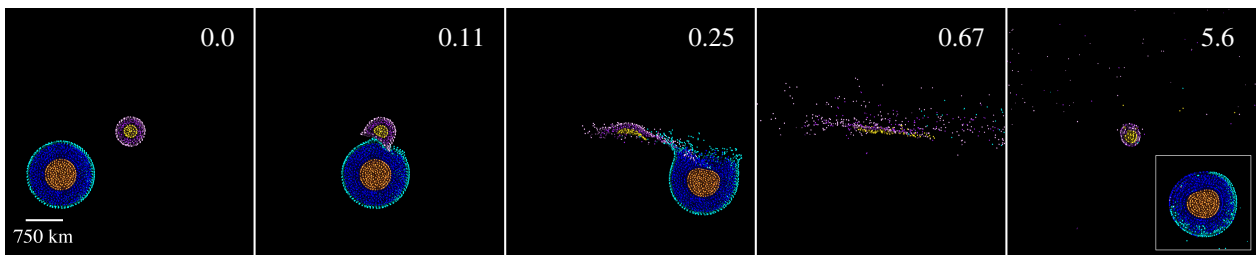
\begin{figure*}
\centering
\vspace{-3mm}

\begin{subfigure}{\textwidth}
\begin{overpic}[width=\textwidth]{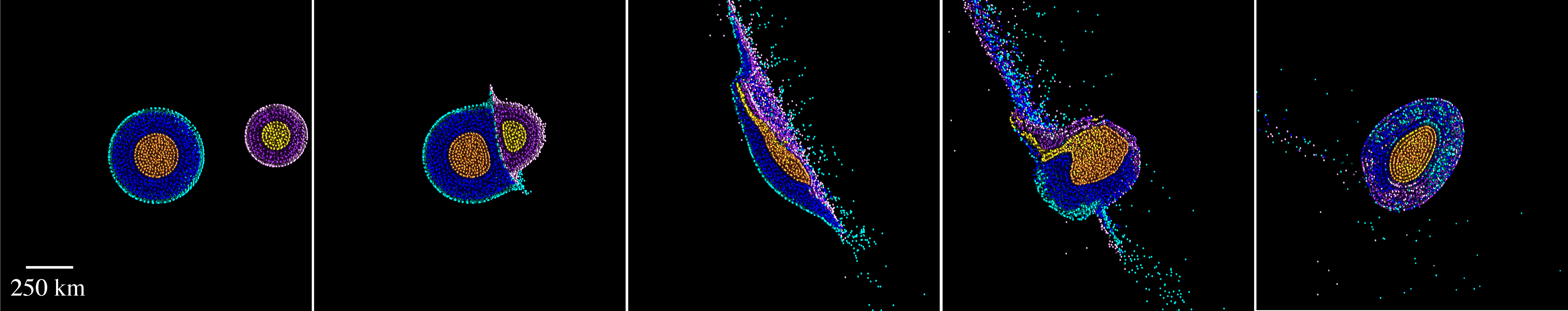}
\put(15.6,16.8){\normalsize \textcolor{white}{0.0}}
\put(34.6,16.8){\normalsize \textcolor{white}{0.22}}
\put(54.6,16.8){\normalsize \textcolor{white}{0.72}}
\put(75.6,16.8){\normalsize \textcolor{white}{1.5}}
\put(95.6,16.8){\normalsize \textcolor{white}{7.0}}
\end{overpic}
\caption{Partial accretion collision between planetesimals with radii of 250\,km and 170\,km (mass ratio 0.3), at an impact velocity of 0.5\,km\,s$^{-1}$ (1.2\,$v_\textrm{esc}$) and impact parameter of 0.25.}
\end{subfigure}

\vspace{3mm}
\begin{subfigure}{\textwidth}
\begin{overpic}[width=\textwidth]{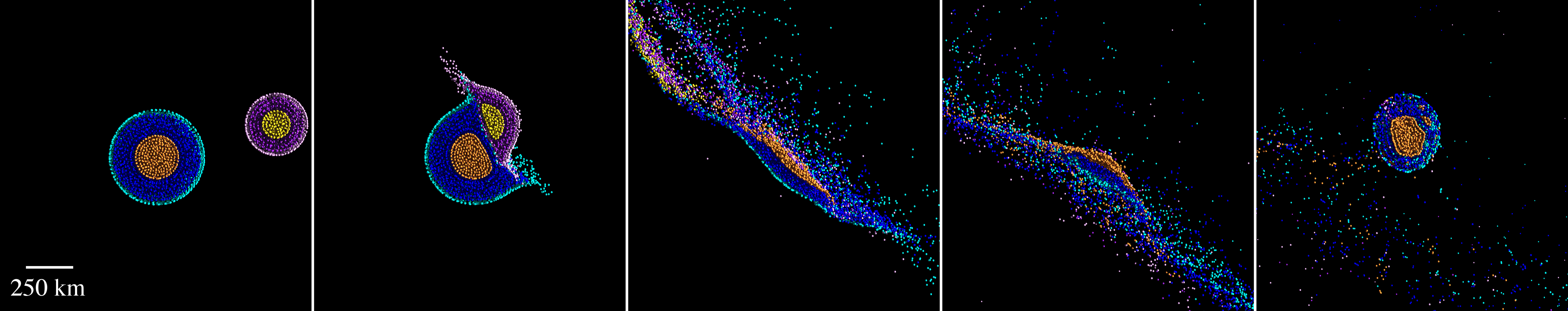}
\put(15.6,16.8){\normalsize \textcolor{white}{0.0}}
\put(34.6,16.8){\normalsize \textcolor{white}{0.08}}
\put(54.6,16.8){\normalsize \textcolor{white}{0.33}}
\put(74.6,16.8){\normalsize \textcolor{white}{1.17}}
\put(94.6,16.8){\normalsize \textcolor{white}{4.67}}
\end{overpic}
\caption{Erosive collision between planetesimals with radii of 250\,km and 170\,km (mass ratio 0.3), at an impact velocity of 1.9\,km\,s$^{-1}$ (4.5\,$v_\textrm{esc}$) and impact parameter of 0.4.}
\end{subfigure}

\vspace{3mm}
\begin{subfigure}{\textwidth}
\begin{overpic}[width=\textwidth]{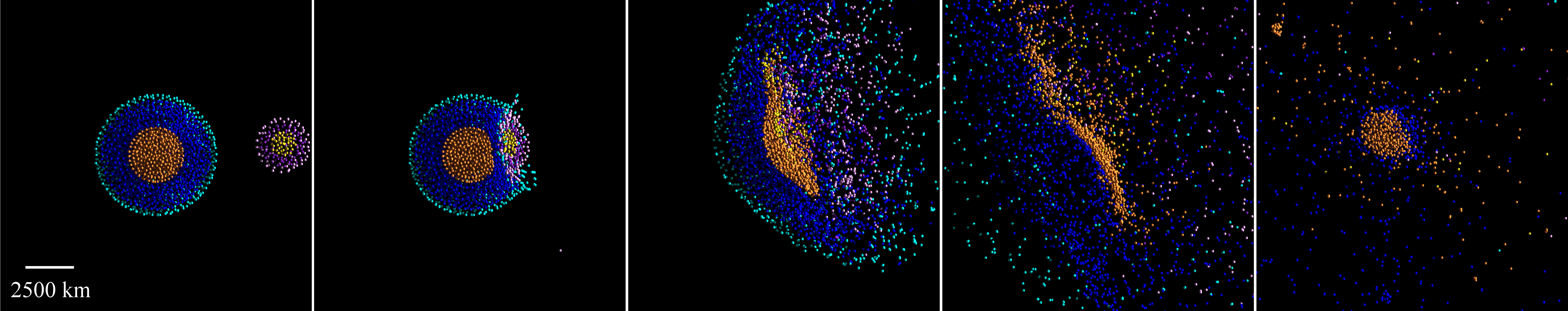}
\put(15.6,16.8){\normalsize \textcolor{white}{0.0}}
\put(34.6,16.8){\normalsize \textcolor{white}{0.05}}
\put(54.6,16.8){\normalsize \textcolor{white}{0.22}}
\put(74.6,16.8){\normalsize \textcolor{white}{0.72}}
\put(95.6,16.8){\normalsize \textcolor{white}{4.0}}
\end{overpic}
\caption{Erosive collision between an embryo with radius of 3160\,km and a planetesimal with radius 1470\,km (mass ratio 0.1), at an impact velocity of 24.1\,km\,s$^{-1}$ (5\,$v_\textrm{esc}$) and impact parameter of 0.1.}
\end{subfigure}

\vspace{3mm}
\begin{subfigure}{\textwidth}
\begin{overpic}[width=\textwidth]{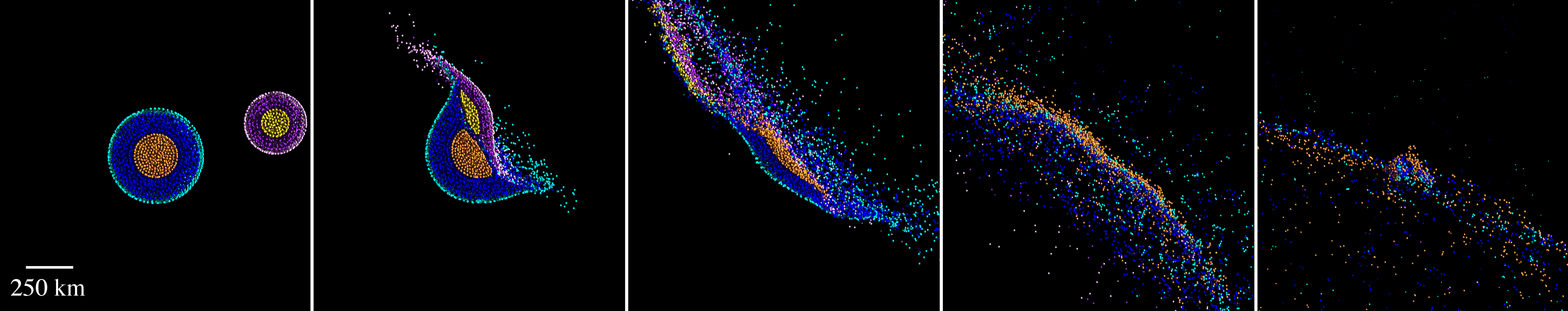}
\put(15.6,16.8){\normalsize \textcolor{white}{0.0}}
\put(34.6,16.8){\normalsize \textcolor{white}{0.06}}
\put(54.6,16.8){\normalsize \textcolor{white}{0.14}}
\put(74.6,16.8){\normalsize \textcolor{white}{0.72}}
\put(95.6,16.8){\normalsize \textcolor{white}{5.0}}
\end{overpic}
\caption{Supercatastrophic disruption due to collision of planetesimals with radii of 250\,km and 170\,km (mass ratio 0.3), at a velocity of 3.8\,km\,s$^{-1}$ (9\,$v_\textrm{esc}$) with an impact parameter of 0.4.}
\end{subfigure}

\vspace{3mm}
\begin{subfigure}{\textwidth}
\begin{overpic}[width=\textwidth]{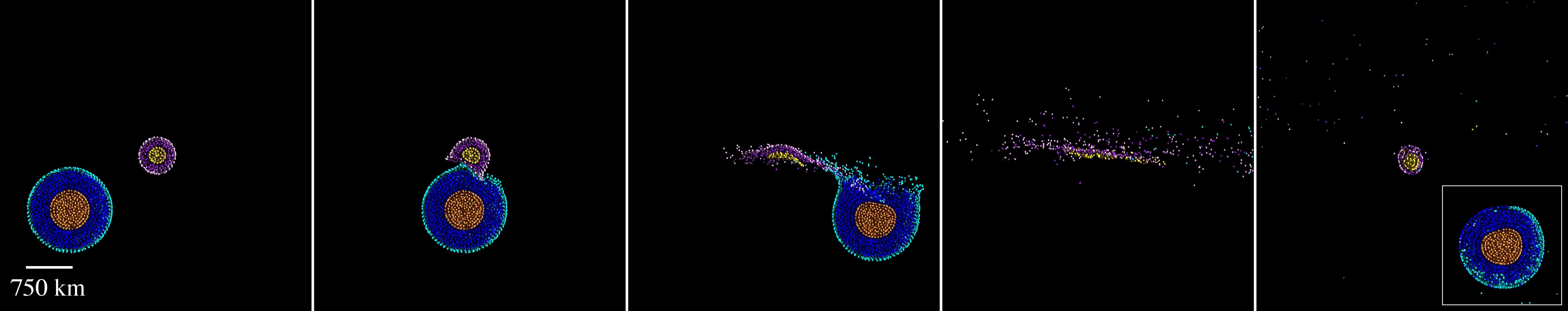}
\put(15.6,16.8){\normalsize \textcolor{white}{0.0}}
\put(34.6,16.8){\normalsize \textcolor{white}{0.11}}
\put(54.6,16.8){\normalsize \textcolor{white}{0.25}}
\put(74.6,16.8){\normalsize \textcolor{white}{0.67}}
\put(95.6,16.8){\normalsize \textcolor{white}{5.6}}
\end{overpic}
\caption{{Hit-and-run collision between planetesimals with radii of 680\,km and 320\,km (mass ratio 0.1), at a velocity of 3.2\,km\,s$^{-1}$ (3\,$v_\textrm{esc}$) with an impact parameter of 0.8. The inset in the last panel shows the remnant of the target.}}
\end{subfigure}

\vspace{-1mm}

\caption{Time series of collisions showing several different outcomes. The numbers indicate the time in hours from the start of the simulation. The target's core, mantle and crust are coloured orange, blue and cyan; the projectile's are coloured yellow, purple and lilac.
Only SPH particles below the equatorial $xy$-plane are shown. 
Each frame is centered on the largest remnant (or second largest in the hit-and-run case). Animations accompanying this figure are available in the supplementary material. 
\label{f:collseq}}
\end{figure*}

The mass fraction of the iron core of the largest remnant as a function of specific impact energy is shown in Figure \ref{f:corefrac}. The dashed curve in this figure is equation 3 from \citet{Marcus09}, scaled by the ratio of the initial core fraction used in this work, 0.22, to the initial core fraction used in \citet{Marcus09}, 0.33. Our results show remarkable consistency with this model despite the lower core fraction.

At low energies the \citet{Marcus09} model is in excellent agreement with our results, however, we find that the core mass fraction generally reaches a plateau at a value of approximately 0.4. 
The second collision sequence shown in Figure \ref{f:collseq} helps to illustrate this: the mantle (and crust) on the lower left of the target's core largely remain behind the remaining core, later becoming a major component of the remnant's mantle.

\begin{figure}
\centering
\includegraphics[width=0.48\textwidth]{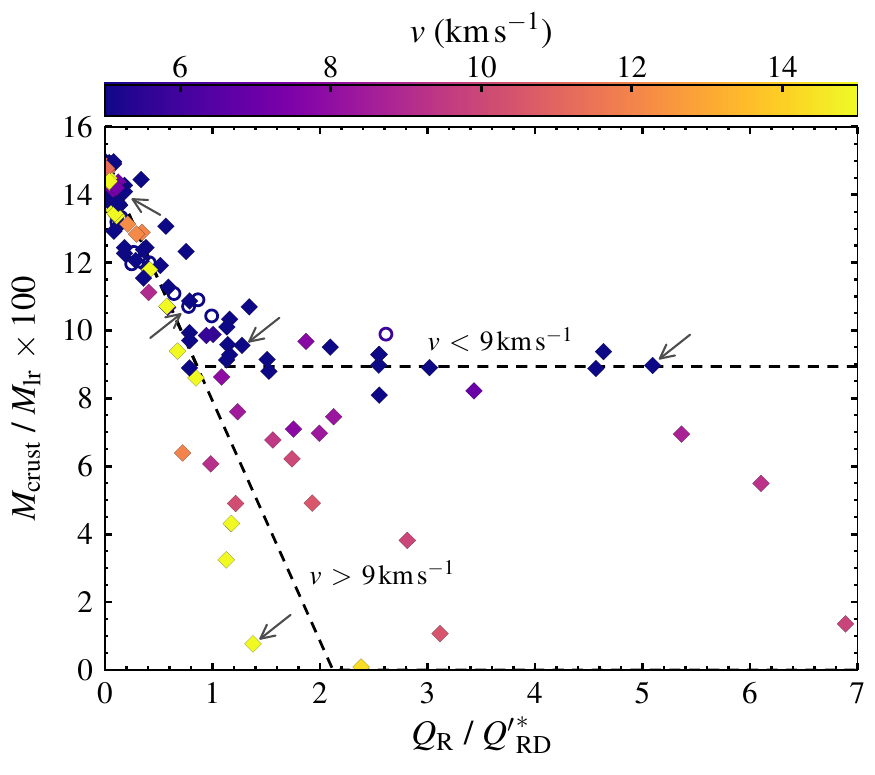}
\caption{Crust mass fraction for the largest remnant vs. normalised specific impact energy. Points are coloured according to impact velocity, {diamonds represent non-grazing collisions and open circles indicate hit-and-runs collisions}. The dashed line corresponds to equation (\ref{e:crust}). The arrows indicate collisions shown in detail in Figure \ref{f:collseq}. {For hit-and-run collisions the crust fraction of the second largest remnant and $Q^\dag_\textrm{R} / Q^{\dag \prime *}_{\mathrm{RD}}$ are plotted.}\label{f:crustfrac}}
\end{figure}

The crust mass fraction of the largest remnant as a function of scaled specific impact energy is shown in Figure \ref{f:crustfrac}. Since the increasing core fraction (Figure \ref{f:corefrac}) requires a preferential loss of mantle in collisions \citep[see also][]{Marcus09,Bonsor15,Carter15}, it is unsurprising that we also find a preferential loss of crust.
Again we see that in many cases the crust mass fraction levels out above specific impact energies of $\sim Q^{\prime *}_{\mathrm{RD}}$ 
as the antipode crust is `protected' along with the mantle. 

Figure \ref{f:crustfrac} reveals that higher velocity impacts lead to lower crust fractions; this can also be seen by comparing the low and high velocity erosive collisions shown in Figure \ref{f:collseq}. 
We find a good fit to the crust fraction of the largest remnant, $f_{\mathrm{crust,lr}}$, with linear relations, dividing the two regimes at a velocity of 9 km$\,$s$^{-1}$, such that
\begin{equation}\label{e:crust}
f_{\mathrm{crust,lr}} = 
\begin{cases}
f_{\mathrm{crust,i}} \left( 1 - 0.47 \left[ \frac{Q_{\mathrm{R}}}{Q^{\prime *}_{\mathrm{RD}}} \right] \right) 	& \text{if}\, \frac{Q_{\mathrm{R}}}{Q^{\prime *}_{\mathrm{RD}}} < 0.85 \\
																			& \text{or}\, v > 9 \,\mathrm{km}\,\mathrm{s}^{-1},\\
0.60\,f_{\mathrm{crust,i}}															& \text{otherwise},\\
\end{cases}
\end{equation}
where $f_{\mathrm{crust,i}}$ is the initial crust mass fraction.

Similarly, the core fraction of the largest remnant can be expressed as
\begin{equation}\label{e:core}
\frac{M_{\mathrm{Fe}}}{M_{\mathrm{lr}}} =
\begin{cases}
f_{\mathrm{Fe,i}} \left( 1 + 0.76 \left[ \frac{Q_{\mathrm{R}}}{Q^{\prime *}_{\mathrm{RD}}} \right]^{1.65} \right), &     \text{if}\, \frac{Q_{\mathrm{R}}}{Q^{\prime *}_{\mathrm{RD}}} < 1.04 \\
																			& \text{or}\, v > 9 \,\mathrm{km}\,\mathrm{s}^{-1},\\
1.82\,f_{\mathrm{Fe,i}}															& \text{otherwise},\\
\end{cases}
\end{equation}
where $f_{\mathrm{Fe,i}}$ is the initial core mass fraction.


\subsection{Crust stripping throughout accretion}\label{s:postprocess}

We have used the crust stripping law determined in the previous section to compare initial compositions with chondritic ratios of refractory lithophile elements to that which would be predicted by our dynamical models. {We applied the crust stripping model
to each collision in a set of $N$-body simulations that represent the intermediate stages of terrestrial planet formation \citep{Carter15}, and thus obtain the bulk silicate abundance of Nd, Sm, Lu, Hf, Th and U in the resulting planetary embryos (Mars mass and above).}

\begin{table*}
\caption{Abundances and bulk partition coefficients, $D$, used for composition calculations, and resulting average embryo compositions (in scenario 2). Bulk silicate Vesta (BSV) abundances based on CI chondrite data from \citet{Palme+Jones03}, partition coefficients taken from \citet{Workman+Hart05}, eucrite composition from \citet{Blicherttoft02} and \citet{Barrat00}, and {KREEP-rich composition based on a combination of an anorthosite outer layer \citet{Munker10} with a KREEP layer \citet{Warren89}, see section \ref{s:crustcomp}. Details of the isotope ratio calculations are given in \ref{a:isotope}}.
\label{t:part}}
\footnotesize
\centering

\begin{tabular}{l r r r r r r r r r r}
\hline
Element	& BSV	& Eucrite & {KREEP} & $D^\textrm{crystal/melt}$ & \multicolumn{6}{c}{Mean embryo abundance relative to BSV} \\
 & (ppm) & (ppm) & (ppm) &  & \multicolumn{2}{c}{model} & \multicolumn{2}{c}{eucritic} & \multicolumn{2}{c}{{KREEP}}\\
  & & & & & calm & GT & calm & GT & calm & GT\\
\hline
Nd	& 1.422 & 5.999	& 13.58 & 0.031 & 0.8148 & 0.7469 & 0.8243 & 0.7593 & 0.8996 & 0.8576 \\ 
Sm	& 0.462	& 1.941 & 3.667 & 0.045 & 0.8235 & 0.7583 & 0.8250 & 0.7602 & 0.9164 & 0.8795 \\ 
Lu	& 0.076 & 0.287	& 0.380 & 0.120 & 0.8592 & 0.8048 & 0.8431 & 0.7838 & 0.9473 & 0.9198 \\ 
Hf	& 0.321 & 1.408	& 2.876 & 0.035 & 0.8174 & 0.7502 & 0.8173 & 0.7502 & 0.9057 & 0.8656 \\ 
Th	& 0.089 & 0.377	& 1.638 &  0.001 & 0.7927 & 0.7181 & 0.8244 & 0.7594 & 0.8078 & 0.7377 \\ 
U	& 0.023 & 0.103	& 0.454 & 0.0011 & 0.7928 & 0.7182 & 0.8167 & 0.7494 & 0.7965 & 0.7230 \\ 
$\epsilon^{143}$Nd & - & - & - & - & 1.25 & 1.77 & 0.10 & 0.14 & 2.17 & 2.97 \\
$\epsilon^{176}$Hf & - & - & - & - & 5.41 & 7.69 & 3.33 & 4.73 & 4.85 & 6.63 \\
\hline
\end{tabular}

\end{table*}
\begin{figure}
\centering
\includegraphics[width=0.48\textwidth]{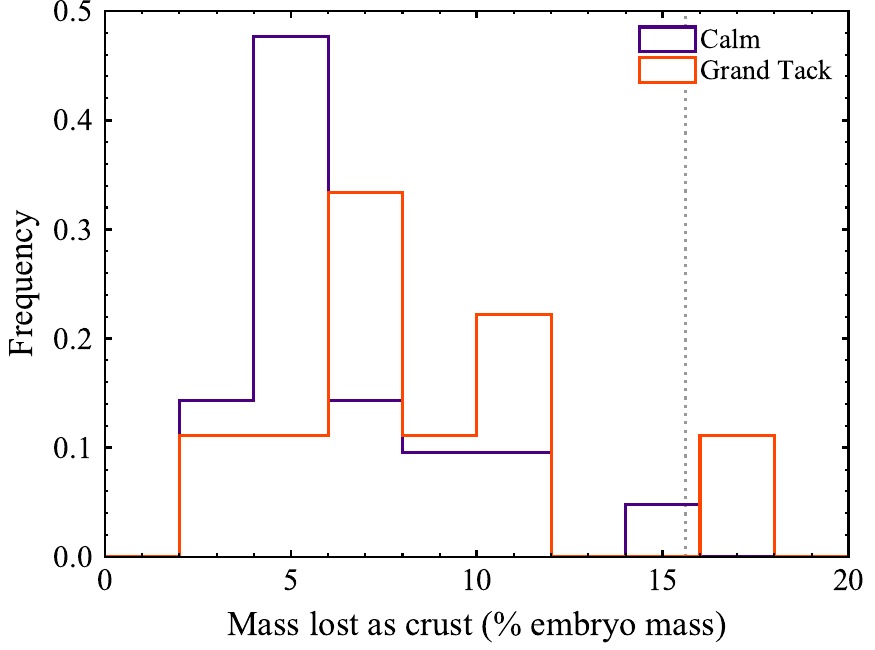}
\caption{Distribution of crust lost during accretion as a fraction of final embryo mass (for scenario 1, in which crust is replaced after each impact). The 20\% melt we have used produces a crust $\sim$15\% of the body mass, as indicated by the dotted line.\label{f:cruststrip}}
\end{figure}
The total mass of crust removed during the accretion of planetary embryos is shown in Figure \ref{f:cruststrip}. Here we only show results for scenario 1, where crust is assumed to be replaced after each impact and so loss is at a maximum.  For scenario two the mass of crust lost is $\sim$2--3\% lower, however, embryos with the lowest mass loss experience approximately the same mass loss in either scenario. The majority of the collisions during the growth of embryos are either low energy accretional or hit-and-run impacts \citep{Bonsor15,Carter15}, so it is not surprising that planetesimals generally lose crust equivalent to only a third to a half of their initial crust mass (Figure \ref{f:cruststrip}). In some cases, particularly for the dynamically hot Grand Tack scenario, growing embryos can lose a mass of crust equivalent to their initial crust. It is important to note that this calculation assumes no reaccretion of stripped crust, and so geochemical calculations based on this loss represent the maximum change.

\begin{figure}
\centering
\includegraphics[width=0.48\textwidth]{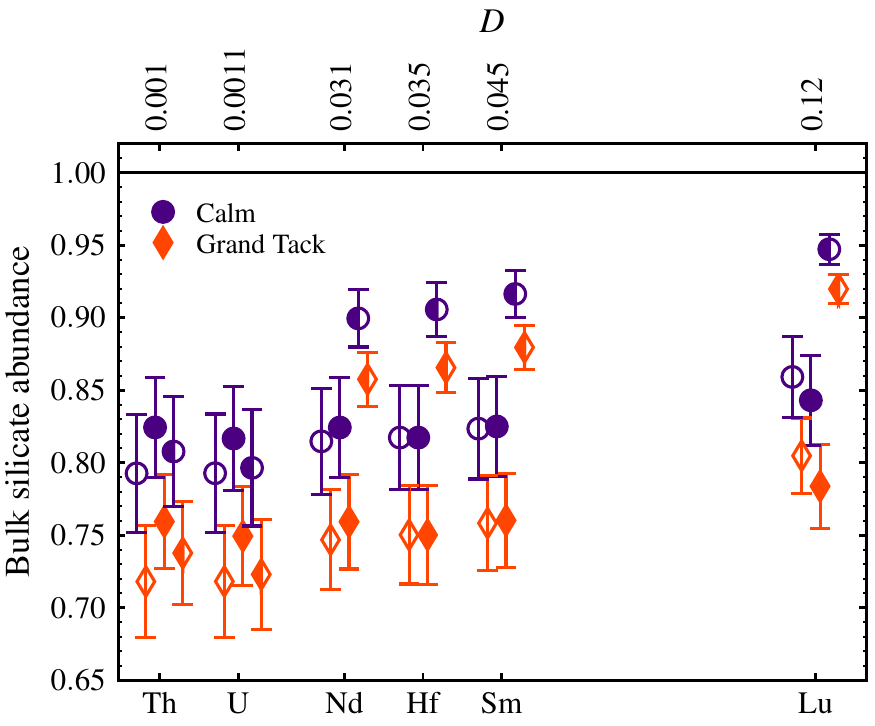}
\caption{Elemental abundances of collisionally processed embryos relative to initial compositions averaged across high resolution simulations from \citet{Carter15} for both calm and Grand Tack simulations. The results for the eucritic crust (filled symbols), model crust (open symbols) and {KREEP-rich crust (half-filled symbols)} are shown (for scenario 2). The elements are not placed on a strict scale, rather they are spaced to reflect their incompatibilities (labelled by their bulk partition coefficients on the upper axis). The symbols represent the mean values, and the ranges indicate the 2$\sigma$ width of the distribution. Note that the depletions for any particular embryo are correlated, such that an object showing the greatest depletion in one element will also display the greatest depletion in the other elements. 
\label{f:abundance}}
\end{figure}
\begin{figure}
\centering
\includegraphics[width=0.48\textwidth]{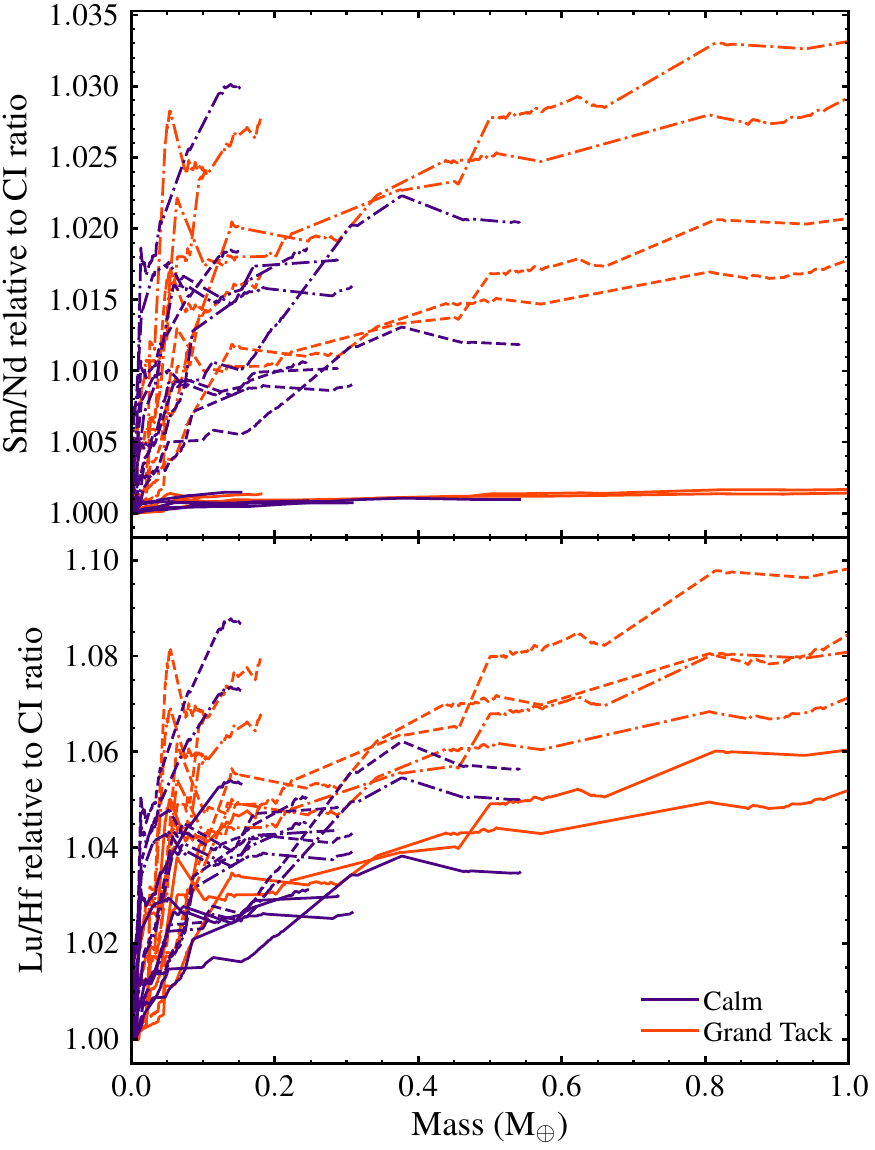}
\caption{Evolution of Sm/Nd and Lu/Hf ratios of individual collisionally processed embryos for example calm and Grand Tack simulations from \citet{Carter15}. All embryos are shown at the same point in time, at which some have grown more massive than others. In both cases the fractionation is lower with a eucritic crust (solid lines) than with the model crust (dashed lines) {or KREEP-rich crust (dot-dashed lines)}. Other simulations show similar results.
\label{f:ratios}}
\end{figure}
The bulk abundance of each element in the final planetary embryos compared to the initial composition is shown in Figure \ref{f:abundance}. Note that ranges shown reflect the variation across the sample of embryos, and the values for any particular embryo are correlated, such that an object showing the greatest depletion in one element will also display the greatest depletion in the other elements. There is little difference between the two scenarios for crustal loss we have tested, and so we show here only the result of the second scenario, in which no new crust is formed 
(a comparison between the two is available in the supplementary material, Figure S1). The elemental depletions do not correspond exactly to the mass of crust lost (Figure \ref{f:cruststrip}) as it is only loss of crust relative to mantle that controls the compositions considered here.

Assuming no reaccretion of lost crustal material, some 20\% to 30\% of the initial inventory of Th and U (key heat-producing elements) can be lost as a result of collisional accretion (Figure \ref{f:abundance}). The influence of accretional erosion on refractory element ratios is illustrated in more detail in Figure \ref{f:ratios}, which shows the evolution of the Sm/Nd and Lu/Hf ratios in planetary embryos. The choice of crustal composition has an important influence on the final Sm-Nd ratio in particular.  This ratio is little changed for loss of eucritic crust, but is significantly perturbed for our model crust, where fractionation is controlled by mantle melting (principally by the partition coefficients of clinopyroxene) and also in our {KREEP-rich crust}.  In a dynamically excited disk (Grand Tack model), an increase in the Sm/Nd ratio of up to 3\% can be achieved. In all cases, Lu-Hf discernibly increases, and again the change is more marked for loss of our model or {KREEP-rich crusts}.


\section{Discussion}\label{s:discussion}

\subsection{SPH impact simulations}

Both the core fraction and the crust fraction of the largest remnant split into two regimes above a specific impact energy approximately equal to $Q^{\prime *}_{\mathrm{RD}}$ (see Figures \ref{f:corefrac} and \ref{f:crustfrac}). 
For lower velocity impacts, the antipode does not reach escape velocity (at least for the combinations of impact geometry and energy explored), thus the mantle and crust near the antipode remain to be incorporated into the largest remnant. In disruptive impacts with velocities above about 9\,km$\,$s$^{-1}$, however, material near the antipode {reaches sufficiently high velocities to be ejected} (see Figure \ref{f:collseq}), and all of the original crust can be removed in high energy impacts. 
The iron fraction data from \citet{Marcus09} show only the upper branch in which a higher mantle fraction is lost. The larger embryos and Earth-mass planets used in those simulations required rather higher impact velocities to reach specific energies above $Q^{\prime *}_{\mathrm{RD}}$, and as such all fell into the antipode-ejection regime.

Close examination of Figures \ref{f:corefrac} and \ref{f:crustfrac} suggests a transition region between the two regimes in a narrow velocity range between 9 and $\sim$12\,km$\,$s$^{-1}$. For the crust mass fraction (in the region between the two lines in the lower right part of Figure \ref{f:crustfrac}) this transition appears to be well ordered with impact velocity increasing as the lower `branch' is approached. It is worth noting that in the simulations of the intermediate stages of planet formation from \citet{Carter15}, there are few impacts above the critical disruption threshold that reach sufficiently high velocities to enter the antipode-ejection regime. Since larger targets are unlikely to experience low velocity collisions that are capable of causing significant erosion, they are unlikely to fall into the plateau regime, and we expect neglecting the transition region and using only two separate branches to have a negligible effect.

{Hit-and-run impacts match well to the `universal law' (dashed line in Figure \ref{f:mlr}, \citealt{Leinhardt12}), and both the core and crust mass fraction scaling laws when the reverse impact onto the projectile is considered (see Figures \ref{f:mlr}, \ref{f:corefrac} and \ref{f:crustfrac}). The target is almost unaffected in hit-and-run collisions (see Figure \ref{f:collseq}), its composition does not change significantly in any of the simulations, though there is a small amount of erosion of the target for higher velocity hit-and-run impacts (see Table S1). The remnant of the projectile generally shows slightly more scatter than the largest remnant of non-grazing impacts, but given the complexities of this regime the agreement is remarkable. It is worth noting, however, that the hit-and-run regime is less well explored than the non-grazing regime in this set of simulations.}

\citet{Genda15} identified a resolution dependence for SPH simulations of disruptive collisions, and suggested that it was necessary to use at least $5 \times 10^{6}$ particles in the target to obtain the value of the catastrophic disruption energy to an acceptable accuracy. Both the standard and higher resolutions used in this work ($2 \times 10^{4}$ or $2 \times 10^{5}$ particles) are substantially lower than \citeauthor{Genda15}'s, however, the 10 times higher resolution checks show at most a 10\% difference in $M_{\mathrm{lr}}$ (see Table S1), well within the scatter shown in Figure \ref{f:mlr}. It is worth noting that our results do show the same lower $Q^{*}_{\mathrm{RD}}$ at higher numerical resolution trend as seen by \citet{Genda15}, but our resolution tests demonstrate that our lower resolution is acceptably accurate.

\subsection{Geochemical consequences of crust stripping}

The chemical effects of crustal stripping shown in section \ref{s:postprocess} give realistic estimates of the {maximum} depletion in incompatible elements on the Earth. Unsurprisingly the dynamically hot Grand Tack scenario leads to a greater depletion of incompatible elements in planetary embryos. Figure \ref{f:ratios} shows that changes in composition due to crust stripping are greatest when the embryos are small, with a large fraction of the loss of incompatible species occurring before the growing embryos reach Mars mass. Smaller bodies are easier to disrupt, and more likely to experience high specific energy collisions, thus crust (and mantle) are more easily removed from small planetesimals than larger embryos. It is expected that the giant impact stage will be dominated by accretion and hit-and-run events, with very few erosive impacts \citep{Stewart12}. This suggests that crust stripping is most important at earlier times, and that the final stages of evolution to fully formed planets will have a smaller effect on incompatible element compositions. 

There is one significant simplification in the post processing model that should be considered: the lack of reaccretion of stripped crust. We have seen previously that reaccretion of eroded mantle and core material is important for the bulk composition of terrestrial planets \citep{Carter15}. 
In \citeauthor{Carter15} mantle and core were tracked directly in the simulations, as was the composition of reaccreted material; whilst here we have post-processed the collisions to estimate the crust loss, a method which cannot track the composition for fragments beyond the second largest remnant. This precludes a direct comparison between the mantle stripping results from \citet{Carter15} and the crust stripping estimated here. 

As crust represents a very small fraction of the total mass it is possible that this material might be removed from the terrestrial planet region without substantial reaccretion. {Whilst the nebular gas remains in the system small debris particles may be lost by spiralling into the star due to aerodynamic drag \citep{Adachi76}, but could be reaccreted by growing planetary bodies before they are removed form the planet forming region. After the gas has been expelled, debris might grind down to particles small enough to be removed by photon pressure \citep[e.g.][]{Kobayashi10}, but it is unclear if this mechanism is efficient whilst accretion is ongoing. The fate of ejecta during planet formation is an important topic that requires further study.} Nevertheless, we would expect at least some ejected crust material to be reaccreted, likely leading to a greater variation in compositions, but also a reduction in the average elemental depletions. {This would work to reduce the compositional changes from the maximum values given in Figures \ref{f:abundance} and \ref{f:ratios}}, but we do not anticipate complete reaccretion onto the originating body as this would be inconsistent with the variation in iron content of planetary embryos seen in \citet{Carter15}.

\citet{Boujibar15} suggested that the elemental disparity but isotopic similarity between the Earth and enstatite chondrites could be explained by collisional erosion of a Si-rich proto-crust. In order to match the high Mg/Si ratio of Earth's primitive upper mantle, they estimated that crust equivalent to $>$15\% of the final planet mass would need to be removed. They also identify that this process must occur early as the Si enrichment of crust-forming melts decreases with increasing pressure (body size). 
The results of our simulations and post-processing suggest that proto-crust can be eroded by collisions during accretion, but the average mass of crust that is removed is a smaller fraction of the total planetary mass (6--9\%) than required by \citet{Boujibar15}. The true discrepancy is larger since we assume a crustal melt fraction of 20\% as opposed to the low degrees of melting (5-7\%) needed by \citet{Boujibar15} and this increases the mass of crustal loss in our scenario. Moreover, {differentiation of planetary bodies occurs early whereas the melt compositions used by \citet{Boujibar15} are based on a primitive enstatite chondrite starting composition. Melt compositions from highly reduced parent bodies that are present in the meteorite record, aubrites, are less Si-rich than the experimental melts of \citet{Boujibar15}. Such empirical crustal compositions would require a greater amount of crustal stripping to produce a terrestrial Mg/Si from an enstatite chondrite starting composition, and we think it unlikely that collisional erosion of crust can account for this. 
However, collisions may play an important role in this regard if they cause vapour loss that results in compositional modification \citep{Pringle14,Hin17}.}

\citet{ONeill08} suggested that loss of an incompatible element enriched crust during accretion could leave Earth enriched in Sm realtive to Nd, and lead to a superchondritic $^{142}$Nd/$^{144}$Nd for bulk Earth. Figure \ref{f:ratios} reveals that using our model or KREEP-rich crust, accretional erosion can naturally lead to a Sm/Nd ratio several percent higher than the chondritic ratio. Since the evolution of planetary embryos considered here is expected to occur early (in the first few Myr of solar system evolution), well before $^{146}$Sm is extinct, 
it is expected that the elevated Sm/Nd ratio would alter the $^{142}$Nd/$^{144}$Nd ratio of the resulting planets. An increase in Sm/Nd of 2\% can result in a planetary embryo with a $^{142}$Nd excess of some 5ppm, or about half the difference between the Earth and enstatite chondrites \citep{Burkhardt16}. We note that the degree of depletion of Nd relative to Sm in our model crust depends on the partition coefficients and degree of melting used in the calculation. The $\sim$2\% depletion of Nd relative to Sm using our model crust shown in Figure \ref{f:ratios} is based on conservative bulk partition coefficients using a composition in which partitioning is controlled by clinopyroxene, appropriate for small bodies with mantle interiors with pressures $<2$\,GPa. Using more extreme ratios, for example those expected if a significant amount of garnet is residual in the source rock, increases this relative Nd depletion. Indeed, using a KREEP-rich crust composition, embryos evolve with a $\sim$4\%  depletion of Nd relative to Sm (Figure \ref{f:ratios}) which can account for the full difference in $^{142}$Nd/$^{144}$Nd between the Earth and enstatite chondrites \citep{Burkhardt16}.

\citet{ONeill08} performed a simplistic version of the crust stripping calculation, constraining the amount of crust lost by the 6\% increase in the Sm/Nd ratio {perceived to be required to produce the 20\,ppm $^{142}$Nd excess reported by \citet{BoyetandCarlson05} in the original exposition of this problem.} They found that this required 54\% loss of a proto-crust formed by a low degree of melting, 2.6\%. Such a low value was rationalised in relation to current day crustal fractions on Earth, but is rather small for typical estimates of planetesimal crusts.  Given the discussion below, however, we have not investigated in further detail the parameter space controlling the exact magnitude of the values produced by loss of our `model' crust.

Recent measurements have demonstrated that the Earth's $^{142}$Nd anomaly is likely nucleosynthetic in origin \citep{Burkhardt16, Bouvier16}, and so a chondritic Sm/Nd for the Earth is again preferred, as has previously been argued \citep{Huang13}. Near chondritic Sm/Nd is reproduced in our simulations if the proto-crust has a composition similar to eucrites, for which we find a small change in Sm/Nd from collisional erosion to a value less than 0.5\% above the chondritic value. This small increase in the Sm/Nd ratio would be consistent with the current interpretation of terrestrial $^{142}$Nd.

\begin{figure}
\centering
\includegraphics[width=0.48\textwidth]{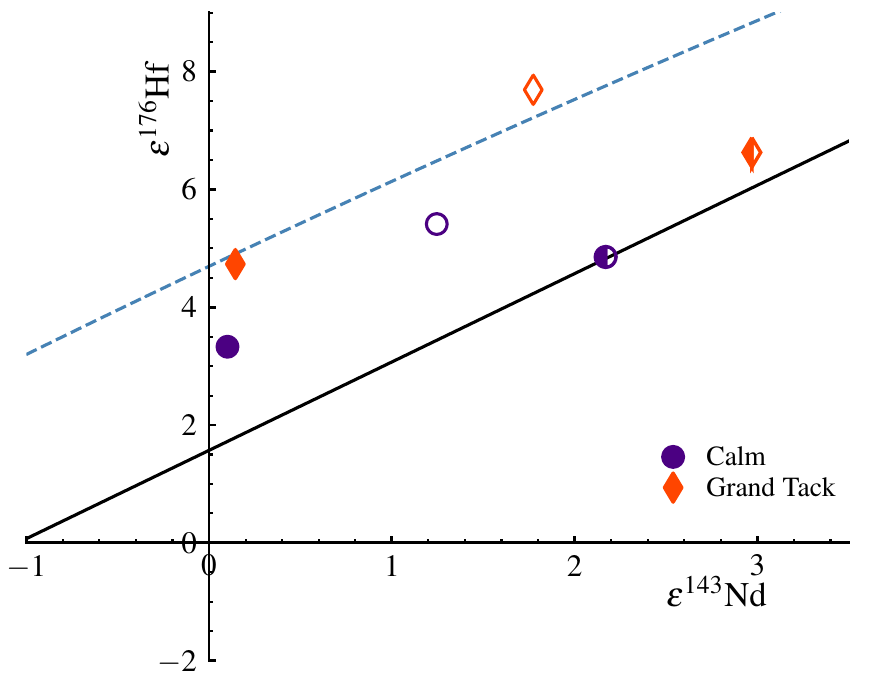}
\caption{Nd-Hf systematics of the Earth and collisionally processed embryos. The solid black line represents the best fit terrestrial array from \citet{Vervoort11}; 
the blue dashed line represents a continental crust-depleted mantle mixing array.  The former was assumed to have $\epsilon^{176}$Hf=-13 and $\epsilon^{143}$Nd=-10 (taken from sedimentary analyses from \citealt{Vervoort11}) and [Nd]=20 and [Hf]=3.7 \citep{Rudnick95} while the latter was taken to have $\epsilon^{176}$Hf=16.8 and $\epsilon^{143}$Nd=9.8 with [Nd]=0.581 and [Hf]=0.157 \citep{Workman+Hart05}. In the absence of a hidden reservoir, values for bulk silicate earth are expected to fall somewhere within these arrays, as is the case for our calculated planetary embryo compositions. Filled symbols represent bulk Earth after stripping of a eucritic crust, open symbols correspond to stripping of our model crust, {and half-filled symbols correspond to stripping of a KREEP-rich crust}.\label{f:Hf-Nd}}
\end{figure}

A longer standing concern over apparently non-chondritic refractory lithophile element abundances in bulk Earth has stemmed from coupled $^{143}$Nd/$^{144}$Nd and $^{176}$Hf/$^{177}$Hf measurements. The array of mantle values for these combined isotopic systems does not pass through the `chondritic cross-hair'  \citep[$\epsilon^{143}$Nd,\,$\epsilon^{176}$Hf = 0,\,0;][where the $\epsilon$ notation indicates parts per ten thousand variation of the $^{143}$Nd/$^{144}$Nd and $^{176}$Hf/$^{177}$Hf ratios respectively from a chondritic reference]{Blicherttoft97}, see Figure \ref{f:Hf-Nd}.  This may simply reflect that the samples defining the mantle array contain variable contributions of subducted crustal components (e.g.\ oceanic crust and sediments) rather than primitive mantle and so might not be expected to pass through a chondritic composition.  Yet it has subsequently become clear that many crustal samples define a similar array that has radiogenic Hf compositions at chondritic $^{143}$Nd/$^{144}$Nd \citep{Vervoort11}. Moreover, if a reconstituted bulk Earth is calculated by mixing depleted upper mantle, as represented by the source of mid-ocean ridge basalts, with average continental crust, the mixing array is convex upwards and has positive $\epsilon^{176}$Hf at $\epsilon^{143}$Nd = 0 (see Figure \ref{f:Hf-Nd}). Accessible samples of the Earth are therefore not readily combined to reproduce a bulk chondritic composition \citep{Blicherttoft97}.  This led to the suggestion of a hidden reservoir at the base of the mantle \citep{Walter04} as was subsequently used as explanation for elevated terrestrial $^{142}$Nd \citep{Labrosse07}.  Less exotically, the lower continental crust might be a suitable complementary reservoir. Whilst accretional erosion has been invoked to account for elevated $^{142}$Nd, to our knowledge, this process has not been explored for its influence on terrestrial  $^{143}$Nd/$^{144}$Nd and $^{176}$Hf/$^{177}$Hf systematics.

In contrast to the Sm-Nd case, loss of 
eucritic crust produces significant Lu-Hf fractionations.  Thus accretional erosion of eucritic, model or KREEP-rich crust will result in planetary embryos with super-chondritic Lu/Hf (Figure \ref{f:ratios}) that evolve into positive modern $\epsilon^{176}$Hf (deviations in $^{176}$Hf/$^{177}$Hf above chondrite, 
see Figure \ref{f:Hf-Nd} and \ref{a:isotope}). {Predicted embryo compositions largely lie within the bounds of the terrestrial Nd-Hf isotopic crust-mantle array \citep{Vervoort11} and mantle-crust mixing curve Figure \ref{f:Hf-Nd}, although loss of KREEP-rich crust results in an extreme composition both in terms of $\epsilon^{176}$Hf and $\epsilon^{143}$Nd.  Assuming a nucleosynthetic origin for $^{142}$Nd excesses on Earth, we favour a scenario with loss of eucritic crust, as this will not produce associated radiogenic $^{142}$Nd anomalies.  Hence, we suggest that the Nd-Hf systematics of the Earth can be 
generated by
crustal erosion of Vesta-like bodies during accretion.  This implies that the heat producing budget of the Earth is $\sim$20\% depleted relative to its initial chondritic protolith (Figure \ref{f:abundance}).}


\section{Summary}

We have conducted smoothed particle hydrodynamics simulations of collisions between differentiated rocky planetesimals and planetary embryos to examine the stripping of planetary crusts during accretion of embryos.
We found the largest remnant mass to be consistent with the \citet{Leinhardt12} scaling law in this intermediate size regime. These simulations show that crust is preferentially lost during impacts, 
and we have developed a scaling law for crust ejection. 
Using this scaling law and the $N$-body simulations of terrestrial planet formation from \citet{Carter15}, we have estimated the maximum effect of collisions during accretion on incompatible element abundances in planetary embryos. 

We find that collisional erosion of proto-crust can lower the budgets of the heat producing elements by $\sim$20\%. 
However, if reaccretion of lost material is very efficient this would substantially reduce any compositional modifications due to collisional erosion, though some variation between planetary bodies would still be likely.

If the composition of proto-crust is similar to the eucrites, planetary Nd isotope ratios remain chondritic; however, alternative crustal compositions can raise Sm/Nd ratios, leading to non-chondritic $^{142}$Nd/$^{144}$Nd, {which would be inconsistent with recent inferences \citep{Burkhardt16, Bouvier16}.}
We find that the Lu-Hf system is notably fractionated even with a eucrite-like crust, which can result in a superchondritic $^{176}$Hf/$^{177}$Hf with chondritic $^{143}$Nd/$^{144}$Nd, in keeping with the inferred composition of Earth.

We have shown that stripping of planetary crusts during accretion can lead to noticeable changes in bulk composition of lithophile elements, but the fractionation is subtle, and sensitive to the unknown efficiency of reacrretion of the stripped material.

\section*{Acknowledgments}

The authors would like to thank S.~J.~Lock for useful discussions and comments, and R. Carlson and an anonymous reviewer for helpful comments that improved this manuscript. This work was carried out using the computational facilities of the Advanced Computing Research Centre, University of Bristol - http://www.bris.ac.uk/acrc/. PJC, ZML, TE \& MJW acknowledge support from the Natural Environment Research Council (grant numbers: NE/K004778/1, NE/M000419/1). TE also acknowledges support from the European Research Council (grant number: 321209, ISONEB).

\appendix

\section{Crustal compositions}\label{a:crustcomp}

{In the post-processing calculations we use Vesta as a suitable analogue for intermediate sized planetesimals, and consider three different approaches for setting crustal compositions.}

{For the `model' Vestan crust we use a simplified model of crust formation based on petrological models discussed in \citet{Ruzicka97} and \citet{Mandler13}. We assume the crust is formed as a 20\% melt fraction (i.e.\ the crust mass is 20\% of the silicate mass of the body, $\sim$15\% of the total mass), which is appropriate for the estimated fraction of eucritic crust. We use a bulk silicate Vesta composition from \citet{Ruzicka97}, which estimates 3$\times$CI chondrite abundances for the refractory lithophile elements studied here, see Table \ref{t:part}. (Chondritic meteorites provide a reliable reference for the ratios of refractory elements, and as lithophile elements, the influence of core formation on element ratios is also insignificant, even if the absolute abundances of the elements require independent estimation). For calculating the composition of a 20\% melt from bulk silicate Vesta, we use the self-consistent set of bulk partition coefficients reported in \citet{Workman+Hart05} which mimic approximately similar conditions.  The calculations are insensitive to the exact values of partition coefficients used, given their low values for all elements considered relative to the melt fraction.  We also note that the batch partitioning scenario we consider can represent either the residue of magma ocean crystallisation (as in \citealt{Ruzicka97} and \citealt{Mandler13}) or partial mantle melts, as envisioned for Vesta by \citet{Stolper77} or for the terrestrial situation modelled by \citet{Workman+Hart05}.}

{In addition to the `model' crust we use analyses of eucrites as a direct sample of Vestan crust.  Given our focus on Sm-Nd and Lu-Hf, we use the high-precision, isotope dilution dataset from \citet{Blicherttoft02}.  For Th and U concentrations we take the average values from \citet{Barrat00}, renormalized so that the mean [Nd] of these data equal the mean [Nd] of the \citet{Blicherttoft02} data. As for the model crust we assume this empirical `eucritic' crust is 20\% of the silicate mass.}

{The incompatible element enrichment in eucritic samples is modest.  More extreme increases in incompatible element abundances and fractionations of one element to another is evident in the final residual liquid of lunar crystallisation, the so-called KREEP layer. This reflects an endmember scenario for differentiation of a larger body. The high-K KREEP composition of \citet{Warren89}, corresponds to a $\sim$0.4\% residual liquid of lunar magma ocean crystallisation \citep{Snyder92}.  This incompatible element enriched layer does not form the crust itself but in an idealised lunar stratigraphy lies beneath the flotation anorthosite crust. Thus, removal of KREEP will occur with simultaneous removal of this overlying plagioclase unit which contains low abundances of the incompatible elements of interest. We have therefore calculated a `crustal' composition that averages a KREEP layer that is 0.375\% of lunar mass with anorthosite representing 4.66\% of lunar mass \citep{Wieczorek13}.  We use elemental abundances for average high-K KREEP from \citet{Warren89} and for the anorthosite using the `enriched' sample 15415 from \citet{Munker10}. For the latter there are no Th nor U concentration data and we assume that the contents of these highly incompatible elements are negligible. For consistency with lunar crystallisation, we assume this `KREEP-rich' crust represents 5.04\% of a body's silicate mass.}

\section{Isotope calculations}\label{a:isotope}

The deviations in present day $^{142}$Nd/$^{144}$Nd, $^{143}$Nd/$^{144}$Nd and $^{176}$Hf/$^{177}$Hf from a chondritic reference are calculated from our collisionally processed Sm/Nd and Lu/Hf ratios assuming an age of 4.567 Gyr. The present day chondritic ratios of $^{142}$Nd/$^{144}$Nd and $^{146}$Sm/$^{144}$Nd were obtained from \citet{Caro03}, and the present day chondritic ratios of $^{143}$Nd/$^{144}$Nd, $^{176}$Hf/$^{177}$Hf, $^{147}$Sm/$^{144}$Nd, and $^{176}$Lu/$^{177}$Hf were taken from \citet{Bouvier08}. The half lives of $^{147}$Sm, $t_{1/2}=1.06\times10^{11}$\,yr, and $^{176}$Lu, $t_{1/2}=3.71\times10^{10}$\,yr, were taken from \citet{Begemann01} and \citet{Scherer01}.

\section*{Supplementary material}

Table S1: Summary of parameters and results of simulated collisions.

{Table S2: Summary of results for second largest remnant of hit-and-run collisions.}

Figure S1: Comparison of processed embryo abundances between different crust formation scenarios.

Animations to accompany Figure \ref{f:collseq}.

\section*{References}

\bibliographystyle{elsarticle-harv}

\clearpage

\includepdf[pages={-}]{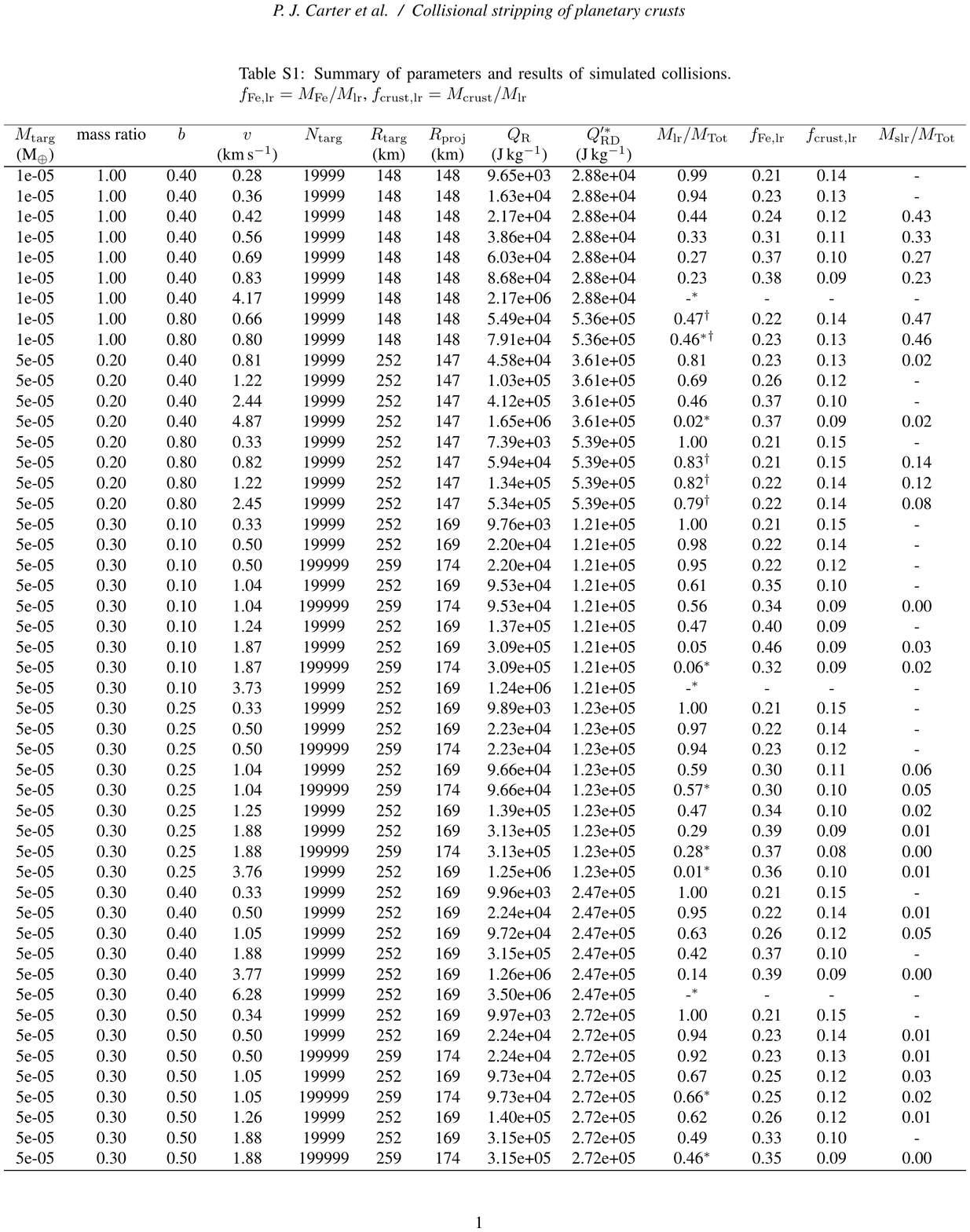}

\end{document}